\documentclass{elsart}
\usepackage{epsf,amssymb}
\journal{Physics Letters A}
\begin{document}
\newcommand{\bra}{\langle}
\newcommand{\ket}{\rangle}
\newcommand{\eq}[2]{\begin{equation}\label{#1} #2 \end{equation}}
\newcommand{\tbf}[1]{{\bf #1}}
\newcommand{\tit}[1]{{\it #1}}
\newcommand{\intRR}{\int\limits_{-\infty}^{\infty}}
\newcommand{\intR}{\int\limits_{0}^{\infty}}
\newcommand{\intRt}{\int_{0}^{\infty}}
\newcommand{\LRD}[1]{\frac{{{\displaystyle\leftrightarrow}
\atop {\displaystyle\partial}}}{\partial #1}}
\newcommand{\lrd}[1]{\stackrel{\displaystyle \leftrightarrow}
{\displaystyle\partial_{#1}}}
\newcommand{\diff}[1]{\partial /{\partial #1}}
\newcommand{\diffm}[2]{\partial#1/\partial #2}
\newcommand{\Diff}[1]{\frac{\partial}{\partial #1}}
\newcommand{\hc}[1]{#1^{\dagger}}
\newcommand{\DiffM}[2]{\frac{\partial #1}{\partial #2}}
\newcommand{\SDiffM}[2]{\frac{\partial^2 #1}{\partial {#2}^2}}
\newcommand{\ds}{\displaystyle}
\newcommand{\sign}{\,{\rm sign}\,}
\textheight=23cm
\begin{frontmatter}
\title{An example of a uniformly accelerated particle detector with non-Unruh response}
\author{A.M. Fedotov\corauthref{cor}},
\corauth[cor]{Corresponding author} \ead{fedotov@cea.ru}
\author{N.B. Narozhny},
\ead{narozhny@theor.mephi.ru}
\author{V.D. Mur},
\address{Moscow Engineering Physics Institute (state university), 115409 Moscow, Russia}
\author{V.A. Belinski},
\ead{belinski@icra.it}
\address{INFN and ICRA, Rome University "La Sapienza", 00185 Rome, Italy and
Institute des Hautes Etudes Scientifiques, F-91440
Bures-Sur-Yvette, France}

\begin{abstract}
We propose a scalar background in Minkowski spacetime imparting
constant proper acceleration to a classical particle. In contrast
to the case of a constant electric field the proposed scalar
potential does not create particle-antiparticle pairs. Therefore
an elementary particle accelerated by such field is a more
appropriate candidate for an "Unruh-detector" than a particle
moving in a constant electric field. We show that the proposed
detector does not reveal the universal thermal response of the
Unruh type.
\end{abstract}

\end{frontmatter}


\section{Introduction}\label{intro}

It was claimed in \cite{Dav,Unruh} that a detector uniformly
accelerated in the empty Minkowski spacetime\footnote{In this
paper we consider a particular case of $1+1$ dimensional
spacetime.} would respond as if it had been placed in a thermal
bath of Fulling-Unruh quanta \cite{Unruh,Fulling} with the
temperature $T_{DU}=a/(2\pi)$, where $a$ is the proper
acceleration of the detector\footnote{We use natural units
$c=\hbar=1$ throughout the paper.}. The temperature $T$ is now
called the Davies-Unruh temperature, while the effect of
"thermalization" of the Minkowski vacuum in accelerated reference
frames is usually referred to as the Unruh effect (more detailed
list of publications concerning this issue can be found in
\cite{PRD}). Let us stress that the most striking feature of this
prediction is {\it universality} of the response of uniformly
accelerating detectors. The latter would mean that this response
were thermal independently of the structure of the given detector
and of the nature of the accelerating force. Such universality, if
existed, would manifest that interaction of accelerated detectors
with the vacuum of a quantum field depended exclusively on general
quantum properties of the vacuum viewed from a uniformly
accelerated reference frame. Moreover, if the effect existed the
uniformly accelerated reference frames would  be physically
preferable among all other reference frames in general relativity.

There exist two different aspects of the Unruh problem \cite{PRD}.
The first one is of purely field theoretical nature and has been
formulated in the original Unruh work \cite{Unruh}. In the
framework of this approach one develops a quantization scheme for
the field restricted to a part of Minkowski spacetime, say Fulling
quantization in the Rindler wedge \cite{Fulling}, and then
attempts to interpret the Minkowski vacuum state in terms of
particle states arising in the process of this quantization, say
Fulling-Unruh particles \cite{Unruh,Fulling}. Another variant of
the field theoretical approach, though physically very close to
the Unruh one, is based on the Bisognano-Wichmann theorem
\cite{Kay,WK}. It is important that the notion of detector in fact
is not exploited by both variants of the quantum field theoretical
approach to the Unruh problem, and thus this approach (if it was
consistent) could give grounds for the aforementioned universality
of the response of a uniformly accelerated detector. Indeed, the
term "Rindler observer" is used in the context of the Unruh
problem exclusively for the sake of convenience and only because
the totality of the world lines of all such "observers", which
however cannot be realized as trajectories of any physical objects
in Minkowski spacetime, completely covers the interior of the
Rindler wedge \cite{PRD}. However, we have shown in
Refs.~\cite{PRD,JETP,PhLet,Ann} that both variants of quantum field
theoretical approach \cite{Unruh,Kay} are physically incorrect,
because they imply that the quantum field satisfies some special
requirement at the boundary of the Rindler manifold which is
incompatible with existence of the Minkowski vacuum state in the
theory. Nevertheless, this conclusion is based on analysis of two
specific sets of arguments \cite{Unruh,Kay} and, strictly
speaking, it does not exclude the possibility of existence of some
more profound evidence in favor of the Unruh effect.

Another aspect of the problem is behavior of physical detectors of
the given structure uniformly accelerating under action  of the
given external force. Certainly, this approach is not so general
as the first one and there is no hope to prove universality of the
Unruh effect in its framework. Therefore this issue is usually
discussed in literature with the purpose of illustration of the
results obtained by means of quantum field theory, see, e.g., the
reviews \cite{Tak,Ros}. However, since there is still no
compelling evidence for the universal behavior attributed to all
uniformly accelerated detectors, the investigation of the response
of particular detectors is of great physical interest. Besides
discussion of possible experimental observations of the effects of
interaction of different quantum systems with external fields,
non-universality of thermal response could be proven within such
approach just by demonstration of at least a single example of a
uniformly accelerated detector which does not reveal the Unruh
behavior.

Consideration of behavior of a uniformly accelerating physical
detector is a rather difficult problem and its treatment in
literature is contradictory. The major difficulty is that a
uniformly accelerating detector (i.e., an elementary particle or a
microscopic bound system) must be considered as a quantum object
moving along a definite classical trajectory. Strictly speaking,
such an assumption is in contradiction with the uncertainty
principle, its range of applicability is very limited and
therefore it must be used with proper care. Since a systematic
relativistic theory of bound states is still absent, different
authors are compelled to use some simplifying assumptions, which
are hard to control. Therefore we will consider below the case of
an elementary particle detector, which, in our opinion, admits the
most consistent analysis at the present moment.

An example of an elementary particle detector accelerated  by a
homogeneous electric field in $1+1$-dimensional spacetime was
considered in Refs.~\cite{PM,Gab}. The considered detector is
described by two equally charged bosons with masses $m$ and $M>m$
interacting with a neutral boson of mass $\mu$. In the absence of
the external electric field there obviously exists transition $M
\to m + \mu$, if $M^2>m^2+\mu^2$. In the presence of the electric
field spontaneous excitation of a $m$-boson into a $M$-particle
accompanied by the emission of $\mu$-particle is also possible. It
was shown in Refs.~\cite{PM,Gab}, that for the special case of
equal accelerations of the charged bosons $(M-m)/m\ll 1$, the
ratio of the transition rates of the two processes $m\to M+\mu$
and $M\to m+\mu$ has the Boltzmann form, with temperature
parameter coinciding with the Davies-Unruh temperature. Though the
latter result holds only for $D=1+1$ spacetime, see Ref.~\cite{NR}, the
authors of Refs.~\cite{PM,Gab} interpret their treatment as purely
quantum mechanical derivation of the Unruh effect since recoil
effects have been taken into account. However the presence of the
electric field violates stability of detector vacuum. Hence the
detector is moving not in empty Minkowski space but in a bath of
pairs of $(m,\overline{m})$ and $(M,\overline{M})$ charged bosons.
Moreover, the authors emphasize the fact that the processes of
detector transitions and the Schwinger process of creation of
boson pairs are in exact equilibrium. Thus relevance of the
suggested detector model to the Unruh effect which implies thermal
equilibrium between detector and vacuum of $\mu$-boson field
remains unclear.

In this paper we consider another mechanism of acceleration of an
elementary particle detector, namely acceleration by a stationary
scalar background. The most important point of our model is
absence of the process of pair creation by the external scalar
field. The detector is described by a fermion $F$ of mass $m$ and
a boson $B$ of mass $M>m$ interacting with a free massless fermion
$\nu$ (which we will call "neutrino"). We consider the
$F$-particle as a ground state, and the $B$-particle as an excited
state of the detector. Just as in Refs.~\cite{PM,Gab} we restrict
our consideration to the $1+1$-dimensional case.

\section{Uniform acceleration by an external scalar field}\label{1}

Let us consider a classical particle in $D=1+1$ spacetime coupled
to an external stationary scalar field $\Lambda(z)$. The Lagrange
function for this particle can be written in the form
$L(z,v)=-m\,e^{\Lambda(z)}\,\sqrt{1-v^2}$, where $v=dz/dt$ denotes
velocity of the particle. The conserved canonical Hamilton
function reads \eq{En_cons}{
H=v\DiffM{L}{v}-L=\frac{m\,e^{\Lambda(z)}}{\sqrt{1-v^2}}=\epsilon=
{\rm const},} and thus for any given value $\epsilon$ of energy
the velocity $v(t)$ can be represented as a function of its
coordinate $z$, $v=v(z)$. The proper acceleration of the particle
can be represented as \eq{accel}{ a=\frac{(dv/dt)}{(1-v^2)^{3/2}}=
\frac{\epsilon}{m}\,\frac{d (e^{-\Lambda(z)})}{dz}.} Hence the
only form of the scalar background $\Lambda(z)$ which can provide
a uniform acceleration of the particle, is $\Lambda(z)=-\ln[|z-z_s|/R]$,
where $R$ is a characteristic length parameter of the background.
The integration constant $z_s$ may be arbitrary due to translation
invariance. It fixes the position of singularity of the external
field, which is unavoidable in the case of a uniformly accelerating
motion.

We choose below $z_s=0$ and consider the particle motion in the
right half of the space with respect to the singularity, $z>0$,
where $\Lambda(z)=-\ln(z/R)$. With such $\Lambda(z)$ one easily
obtains from Eq.(\ref{En_cons}) for particle trajectories the
equation: $z(t)=\sqrt{t^2+a^{-2}},$ where proper acceleration is
equal to $a=\epsilon/(mR)$ and the integration constant is chosen
so that the time coordinate of the turning point of the trajectory
is $t_0 =0.$ It is worth noting that the totality of obtained
trajectories with $0<\epsilon<\infty$ completely fill the interior
of the right (Rindler) wedge of Minkowski spacetime. Therefore
acceleration by the scalar background is more appropriate for
studying the Unruh problem than acceleration by a homogeneous
electric field. In the latter case acceleration of a particle is
determined solely by the field strength and is independent of
initial conditions. The totality of such trajectories can cover
the whole Minkowski spacetime but of course cannot be restricted
to any separate wedge of it.

\section{Elementary particle detector in scalar background}
\label{Quantization}

We will discuss now behavior of a quantum detector accelerated by
the scalar background introduced in the preceding section. As a
rule, a kind of "two-level" system is used as a detector while
discussing the Unruh effect \cite{Tak,Ros,PM,Gab}. Such a detector
is accelerated by an external force which does not affect however
the field responsible for transitions of the "two-level" system
and hence the detector can be considered moving in undisturbed
vacuum of the latter field. Since the scalar background
$\Lambda(z)$ accelerates any massive particle, the field
responsible for transitions of our detector should be massless.
However it is well known, see, e.g., Ref.~\cite{SW}, that in
$D=1+1$ theories a massless scalar particle does not exist even as
a mathematical object. Therefore we have to consider a "two-level"
system interacting with a massless fermion
("neutrino-antineutrino") field as a model of the detector and
hence one of the particles constituting our detector must also be
a fermion. We start with discussion of solutions of
Klein-Fock-Gordon (KFG) and Dirac equations in the presence of the
scalar background $\Lambda(z).$

The squared Hamilton function (\ref{En_cons}) for a boson of mass
$M$ can be written in the form $H^2=\Pi^2+M^2e^{2\Lambda(z)}$,
where $\Pi=\partial L/\partial v$ is momentum of the particle. The
KFG equation is obtained from this relation by the standard
substitutions $H\to i\diff{t}$, $\Pi\to -i\diff{z}$: \eq{KFG}{
\left\{\SDiffM{}{t}-\SDiffM{}{z}+M^2\,e^{2\Lambda(z)}\right\}\Psi_B(t,z)=0.}

Positive frequency ($\epsilon>0$) solutions of Eq.(\ref{KFG}) has
the form  $\Psi_B(t,z)=\psi_{B\epsilon}(z)e^{-i\epsilon t}$ where
the functions $\psi_{B\epsilon}(z)$ satisfy the stationary
Schr\"{o}dinger equation \eq{sp_KFG}
{\psi_{B\epsilon}''(z)+\left(\epsilon^2-\frac{M^2R^2}{z^2}\right)\psi_{B\epsilon}(z)=0.}
The penetrability coefficient for the barrier $M^2R^2/z^2$, which
controls the possibility of quantum tunnelling through the
singularity of the background, can be estimated quasiclassically
(see, e.g., Ref.~\cite{LL}), \eq{D}{ D\sim
\exp\left\{-2\int\limits_0^{1/a}dz\,
\sqrt{\frac{M^2R^2}{z^2}-\epsilon^2}\right\}=0,} and is equal to
zero since the integral in the argument of the exponential in
(\ref{D}) is logarithmically divergent on its lower limit. It
means that the potential barrier which surrounds the singularity
$z=0$ is impenetrable even on the quantum level. As a consequence,
the boundary condition $\psi_{B\epsilon}(0)=0$ should be imposed
at the singularity. The field modes satisfying this boundary
condition read \eq{sol_st}{
\Psi_{B\epsilon}(t,z)=\psi_{B\epsilon}(z)e^{-i\epsilon
t}=\sqrt{\frac{z}{2}}\; J_{\sqrt{M^2R^2+1/4}}(\epsilon
z)\,e^{-i\epsilon t},} where $J_{\kappa}(\epsilon z)$ is the
Bessel function, and the normalization constant is defined by the
condition \eq{inner_pr}{ i\intR dz\;\Psi_{B\epsilon}^*(t,z)\LRD{t}
\Psi_{B\epsilon'}(t,z)=\delta(\epsilon-\epsilon').} The Dirac
equation in the accelerating scalar background reads \eq{Dirac}{
\left[i\left(\gamma_0\Diff{t}-\gamma_1\Diff{z}\right)- m
e^{\Lambda(z)}\right]\Psi_F(t,z)=0,} where $\Psi_F$ is a
two-component function. We adopt the following representation for
the Dirac $\gamma$-matrices, $\gamma_0=\sigma_3$,
$\gamma_1=i\sigma_1$, where $\sigma_i$ are the standard Pauli
matrices. Due to the impenetrability of the potential barrier at
the point $z=0$ (see the Eq.(\ref{D})), one should impose at this
point the boundary condition of vanishing of the local scalar
current $s=\bar\Psi_F\Psi_F$, $s(t,0)=0$. The positive-
(negative-) frequency modes, which obey such a boundary condition
and are normalized by the relations \eq{norm_D}{ \intR
dz\,\Psi_{F\epsilon}^{(\pm)\,\dagger}\Psi_{F\epsilon'}^{(\pm)}=
\delta(\epsilon-\epsilon'),\quad \intR
dz\,\Psi_{F\epsilon}^{(\pm)\,\dagger}\Psi_{F\epsilon'}^{(\mp)}=0,}
read \eq{D_modes}{
\Psi_{F\epsilon}^{(\pm)}(t,z)=\psi_{F\epsilon}(z)e^{\mp i\epsilon
t}=\frac{\sqrt{\epsilon z}}{2}\, \left(\begin{array}{c}\ds
J_{mR-1/2}(\epsilon z)\pm J_{mR+1/2}(\epsilon z)\\
J_{mR-1/2}(\epsilon z)\mp J_{mR+1/2}(\epsilon
z)\end{array}\right)\, e^{\mp i\epsilon t}.} Finally, we write
down the positive- (negative-) frequency solutions for the free
massless Dirac equation  \eq{nu_sol}{
\Psi_{\nu\,p}^{(\pm)}(t,z)=\psi_{\nu p}(z)e^{\mp i|p|
t}=\frac1{2\sqrt{\pi}}\left(
\begin{array}{c} 1\\ \pm i\sign(p)\end{array}\right)\,
e^{i(pz\mp |p|t)},} $ -\infty<p<+\infty$. Since massless particles
do not interact with the scalar background the functions
(\ref{nu_sol}) do not obey any boundary condition at the point
$z=0$. We use the sets of modes
(\ref{sol_st}),(\ref{D_modes}),(\ref{nu_sol}) for quantization of
respectively the massive boson $\Phi_B$, fermion $\Phi_F$ and
neutrino $\Phi_\nu$ fields.

Let the Lagrangian of interaction between the three fields be of
the form $\mathcal{L}_{{\rm
int}}=\lambda(\bar\Phi_F\Phi_{\nu}+\bar\Phi_{\nu}\Phi_F) \Phi_B$.
We will consider, first, the decay of the massive fermion
$F_{\epsilon}\to B_{\epsilon'}+\nu_p$ which does not occur in the
absence of the scalar background. Let the initial state of the
fermion be a wave packet $|F\ket=\int_0^{\infty} dE\,C_i(E)
f_{E}^{\dagger}|{\rm vac}\ket$ with the spectral weight function
$C_i(E)$ normalized by the condition $\int_0^{\infty}
|C_i(E)|^2\,dE=1$ and $f_{E}^{\dagger}$ is fermion creation
operator. Then the matrix element of the decay in the first order
of perturbation theory is given by \eq{Ampl}{
\begin{array}{l}\ds \bra B_{\epsilon'},\nu_p|F\ket=\\ \\ \ds
=-i\lambda\intRR dt\intR dz \intR
dE\,C_i(E)\Psi_{B\,\epsilon'}^*(t,z){\bar\Psi}_{\nu
p}^{(+)}(t,z)\Psi_{FE}^{(+)}(t,z).\end{array}} Substituting
Eqs.(\ref{sol_st}),(\ref{D_modes}),(\ref{nu_sol}) into
(\ref{Ampl}) we obtain \eq{Ampl1}{ \bra B_{\epsilon'},\nu_p|F\ket=
-i\lambda C_i(\epsilon'+|p|)I(\epsilon',p),} where
\eq{I}{I(\epsilon',p)= \intR dz \psi_{B\,\epsilon'}^*(z)
{\bar\psi}_{\nu
p}^{(+)}(z)\psi_{F,\epsilon'+|p|}^{(+)}(z).}In the case of
narrow spectral weight function of the initial packet, i.e. when
all probable values of the energy of the $F$-particle are very
close to $\epsilon$, we can substitute
$|C_i(E)|^2\approx\delta(E-\epsilon)$. Then we get for the
differential probability of the decay \eq{Prob}{
\begin{array}{l}\ds dW_{F\to
B}(\epsilon',p|\epsilon)=\left\vert\bra B_{\epsilon'},\nu_p|F\ket
\right\vert^2\,d\epsilon'\,dp=\\ \\ \ds
=\lambda^2\delta(\epsilon-\epsilon'-|p|) \left\vert
I(\epsilon',p)\right\vert^2 \,d\epsilon'\,dp.
\end{array}}
It is easy to see that the integral (\ref{I}) is divergent at
$p\rightarrow 0.$ Indeed, the asymptotic behavior of the modes
(\ref{sol_st}),(\ref{D_modes}) is given by the following
expressions \eq{asympt}{
\begin{array}{l}\ds
\Psi_{B\epsilon}(t,z)\sim\frac1{\sqrt{\pi\epsilon}}\;
\cos(\epsilon z-\alpha)\,e^{-i\epsilon t},\quad \alpha=
\frac{\pi}2\left(\sqrt{M^2 R^2 +\frac{1}{4}}+\frac12\right),\\ \\
\ds \Psi_{F\epsilon}^{(+)}(t,z)\sim\frac1{\sqrt{\pi}}\,
\left(\begin{array}{c}\ds \cos(\epsilon z-\beta)\\ -\sin(\epsilon
z-\beta)\end{array}\right)\, e^{-i\epsilon t},\quad
\beta=\frac{\pi}{2}\left(mR+\frac12\right).
\end{array}}
Note that the asymptotic expressions (\ref{asympt}) in fact
coincide with $\Psi-$functions of free massless particles. This is
because the mass defect at long distance from the singularity
equals to the mass of the particle and the latter moves almost
with the speed of light. Substituting expressions (\ref{asympt})
into the integral (\ref{I}) at $p=0$, we arrive to a linear
divergency of the integral (\ref{I}) at $z\rightarrow \infty$. It
means that $I(\epsilon',p)\sim 1/p\,, p\rightarrow 0,$ and we get
for the probability (\ref{Prob}) \eq{Prob2}{ dW_{F\to
B}(\epsilon',p|\epsilon) \sim
\frac{\lambda^2}{\epsilon}\delta(\epsilon-\epsilon'-|p|)
\,d\epsilon'\,\frac{dp}{p^2},\quad p\rightarrow 0. } From this
expression it is evident that the total probability of the decay
is divergent. This fact has a simple physical explanation. Indeed,
suppose the decay occurs after reflection from the singularity,
when the wave packet $|F\ket$ is moving to the right. Due to
special properties of the fermion current $\bar\Psi_F\Psi_F$ in
$D=1+1$ dimensions (see, e.g., Ref.~\cite{Sussk}), the neutrino with
energy $|p|$ is emitted strictly to the left. Thus, due to the
energy conservation law, the boson $B$ with the energy
$\epsilon'=\epsilon-|p|$ can be emitted either to the right (in
such a case the momentum transferred to the external field is
$\Delta P=2|p|$) or to the left ($\Delta P=2\epsilon$). It is easy
to see that the second process can occur only at distances
$z\lesssim\epsilon^{-1}$ from the singularity. Hence only the
first process contribute to the total probability at long times
since $p$ can be arbitrary small. If the time of observation
$\tau$ is finite, then the neutrinos with the energies
$|p|\lesssim\tau$ can not be emitted. Thus for the total
probability at long time we obtain \eq{Prob1}{\ds W_{F\to
B}(\epsilon)\sim\frac{\lambda^2}{\epsilon} \int\limits_{1/\tau}^{}
\frac{dp}{p^2}= \frac{\lambda^2}{\epsilon}\,\tau ,\quad
\tau\to\infty.} It means that there exists the decay rate
$\mathcal {R}_{F\to B}\sim\lambda^2/\epsilon$ (more rigorous
derivation of this result will be given in a forthcoming
publication).

The lowest-order amplitude of the process $B_{\epsilon'}\to
F_{\epsilon}+\widetilde{\nu}_p$ reads
\eq{Ampl_i}{\begin{array}{l}\ds \bra
F_{\epsilon},\widetilde{\nu}_p|B\ket=\\
\\ \ds =-i\lambda\intRR dt\intR dz \intR
dE\,C_i(E)\Psi_{B\,E}(t,z)
{\bar\Psi}_{F\epsilon}^{(+)}(t,z)\Psi_{\nu
p}^{(-)}(t,z).\end{array}} It is easy to see that the decay
rate for this process can be calculated by exactly the same method
which have been used for calculation of $\mathcal {R}_{F\to B}$
and give the result $\mathcal {R}_{B\to
F}\sim\lambda^2/\epsilon'$.

\section{Long-time behavior of the accelerated detector}\label{3}

It was shown in the preceding section that at long times $\tau$
($1/\epsilon\ll\tau\ll\epsilon/\lambda^2$) the processes of
neutrino, antineutrino emission are characterized by decay rates.
It means that master equations can be used to study the long-time
behavior of the detector. One can easily obtain from
Eq.(\ref{Prob2}) that the average neutrino energy emitted for time
$\tau$ is of the order ${\bra |p|\ket} \sim
\lambda^2/\epsilon\,\ln\epsilon\tau$. Hence, with regard for the
upper limit for $\tau$ admitted in the framework of perturbation
theory, we have $\bra |p|\ket\ll\epsilon$. This result means that
neutrinos are effectively emitted in zero mode at long times and
thus the master equations for the problem constitute a set of
differential equations rather than of integro-differential ones.

Denote by $n_F$, $n_B$ the average numbers of fermions and bosons
with the energy $\epsilon$ and by $n_{\nu}$, $n_{\widetilde{\nu}}$
the average numbers of neutrinos and antineutrinos in the zero
mode. Let $\mathcal {R}_{F\to B}=\mathcal {R}_{B\to F}\equiv
\mathcal {R}.$ Then the corresponding master equations, which take
into account effects of quantum statistics, read \eq{kinetics2}{
\begin{array}{l}\ds
\frac1{\mathcal
{R}}\frac{dn_F}{dt}=n_B(1-n_F)(1-n_{\widetilde{\nu}})-n_F
(1+n_B)(1-n_{\nu}),
\\ \\ \ds
\frac1{\mathcal {R}}\frac{dn_{\nu}}{dt}=n_F
(1+n_B)(1-n_{\nu}),\quad n_F+n_{\nu}- n_{\widetilde{\nu}}=1,\quad
n_F+n_B=1.
\end{array}}

\begin{figure}[eeeee]
\epsfxsize=15cm \centerline{\epsffile{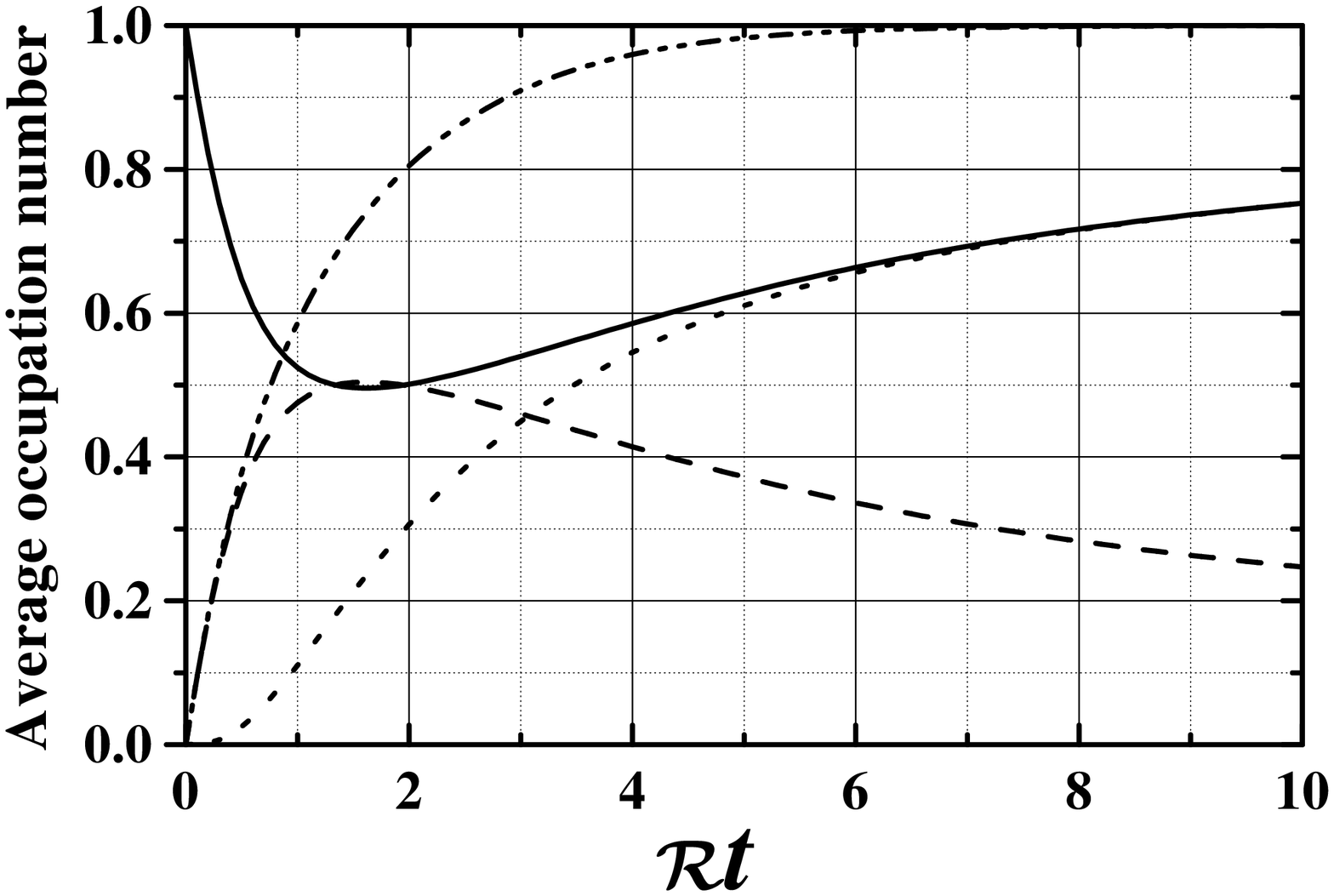}} \caption{
Long-time behavior of the occupation numbers $n_F$ (a solid line),
$n_B$ (a dash line), $n_{\nu}$ (a dash dot dot line), and
$n_{\widetilde\nu}$ (a dot line).}
\label{graph}
\end{figure}

The results of numerical solution for the system (\ref{kinetics2})
are presented in Fig.~\ref{graph}. The initial conditions adopted
for the presented solution are $n_F(0)=1$ and $n_B(0)=n_{\nu}(0)=
n_{\widetilde{\nu}}(0)=0$.

It is seen from Fig.~\ref{graph} that, while $t\ll \mathcal
{R}^{-1}$, the initial fermion decays into boson and neutrino. For
$t\ge \mathcal {R}^{-1}$ the number of neutrinos in the zero mode
tends to unity exponentially fast,
$1-n_{\nu}(t)\propto\sqrt{t}\,e^{-\mathcal {R}t}$. However, when
$n_\nu$ comes to unity, this channel owing to the Pauli principle
becomes closed. The number of antineutrinos is initially growing
much more slowly, since the antineutrinos are created by boson
decay, and no bosons are initially present. After $n_\nu$ comes to
unity, the created boson decays into fermion and antineutrino.
Finally, the number of antineutrinos also comes to unity, and this
channel also becomes closed. Asymptotically the resulting state
contains a fermion and a $\nu\widetilde{\nu}$-pair in the zero
mode. However, this state is achieved very slowly,
$1-n_F(t)\approx 1-n_{\widetilde{\nu}}(t)\approx n_B(t)\approx
(2\mathcal {R}t)^{-1/2}$ as $t\to+\infty$.

Thus the long time behavior of the detector accelerated by scalar
background crucially differs from the behavior one could predict
on the basis of the Unruh conjecture. Indeed, instead of arriving
to some "thermal equilibrium" state, the detector creates a
$\nu\widetilde{\nu}$- pair and returns to its initial state. Note
that, if we change the initial conditions for the detector
$n_F=1,\, n_B=0\, \rightarrow n_F=0,\, n_B=1$,
we will have boson in the final state instead of the fermion.
It means that the detector remembers its initial state and hence one
by no means can talk about equilibrium in the final state of this
particular case of detector. From this we conclude that the thermal
Unruh response is not a universal property of all uniformly accelerated
detectors but can be revealed by detectors of some special nature, see
Refs.~\cite{Ros,PM}.

This work was supported by the Russian Fund for Basic Research
under the projects 00--02--16354 and 01--02--16850, and by the
grunt 1501 of the Ministry of Education of Russian Federation.

\end{document}